# Non-saturation intensity dependence of anisotropic third-order optical nonlinearity approaching the damage threshold in ZnSe and GaP

JIANPENG YE[1] AND MIN HUANG[1,*]

[1]*State Key Laboratory of Optoelectronic Materials and Technologies, School of Physics, Sun Yat-sen University, Guangzhou 510275, People's Republic of China*
*\*syshm@163.com*

**Abstract:** The intensity dependence of anisotropic third-order optical nonlinearity approaching the damage threshold in ZnSe and GaP crystals is studied by the femtosecond laser pump-probe measurements, which can greatly reduce the laser-matter interaction length and thus realize the probing of orientation-dependent characteristics of nonlinear optical phenomena in the near-damage-threshold intensity regime without significant photon depletion. In the measured transient 3D map, the typical third-order nonlinear optical signals of two-beam coupling (TBC) and two-photon absorption (TPA) can be clearly found out, which both exhibit the pronounced orientation-dependent periodic modulation corresponding to a specific lattice symmetry. Interestingly, the further fixed-delay-time measurements focusing on TBC and TPA confirm that the modulation amplitude of the orientation-dependent curves always increases with the increase of pump intensity towards the damage threshold, which has not been observed in previous studies. Such a definite upward trend of orientation-dependent third-order nonlinear optical effects in the near-damage-threshold regime indicate that, as long as the laser-matter interaction length is small enough, the third-order nonlinear optical phenomena can still be in a non-saturation physical regime till the damage threshold, and thus exhibit significant crystallographic dependence as that of laser-induced damage at the similar intensity ranges.

## 1. Introduction

For the crystallographic dependence of nonlinear optical properties of crystals, the linearly-polarized femtosecond (fs) laser-induced effects of nonlinear absorption [1-8], material damage [9-12], high-harmonic generation [13,14], or supercontinuum emission [15] always exhibit significant crystal orientation dependence. For example, the transmittance of a high-intensity linearly-polarized fs laser propagating through a transparent crystal is influenced by the crystal orientation for the orientation-dependent nonlinear absorption [2,3,6,8]; At the near-damage-threshold intensity, the area of surface periodic structures induced by fs laser irradiation strongly depends on the angle relationship between laser polarization and crystal orientation [9-12]; The intensity of linearly-polarized fs laser-induced high-order harmonic generation on crystals seriously depends on the crystal orientation [13,14]; The depolarization characteristics of the supercontinuum spectrum induced by fs laser inside a crystal are significantly affected by the relationship between laser polarization and crystal orientation [15]. Theoretically, these phenomena are all related to the strong crystallographic dependence of nonlinear optical properties of crystals under strong fs-laser irradiation, and they can be explained by the crystal orientation dependence of nonlinear polarizability or strong field ionization rate of a certain crystal.

Because above crystal orientation-dependent phenomena all have the nonlinear optical essence, the fs-laser intensity plays an important role in them. Experimentally, the dependence of the amplitude of periodic modulation on laser intensity can present the fundamental characteristics of these phenomena, which is of great significance for understanding the physical mechanisms of them. In general, as reported by previous studies [2,3,16], with the increase of laser intensity to some extent, the modulation amplitude of crystal orientation-dependent signals often exhibits a similar saturation trend, and even an attenuation trend. However, it is worth noting that for different crystal orientation-dependent phenomena, the intensity ranges corresponding to the signal saturation regime may be inconsistent. Typically, for nonlinear absorption transmission measurements, the modulation amplitude peak of orientation-dependent transmitted laser intensity will appear in a specific intensity range below the laser damage threshold, and as the laser intensity further approaches the damage threshold, the modulation amplitude will decrease [2,3,16]; For the fs laser-induced periodic surface structures, its crystal orientation dependence is most pronounced near the damage threshold, and as the laser intensity further increases and moves away from the damage threshold, the modulation amplitude of orientation-dependent ablation will rapidly decrease [9]. It can be seen that the modulation amplitude peaks of these two crystal orientation-dependent phenomena occur in the different laser intensity ranges, resulting in incompatibility between the occurrence of the two phenomena: the modulation amplitude of orientation-dependent nonlinear absorption approaches zero at the damage-threshold laser intensity, whereas at this intensity the orientation-dependent laser-induced damage is in the most pronounced regime. Considering that the two phenomena have similar physical origins—nonlinear absorption (ionization) under strong laser fields, the incompatible dominant laser intensity ranges between them seems to be contradictory. The physical origin of this issue needs further experimental and theoretical research to solve.

For the abovementioned crystal orientation-dependent phenomena, in addition to the core physical quantity of laser intensity, the significant impact of the laser-matter interaction length on these phenomena is easily ignored. As the previous study revealed [16], when the laser-matter interaction length $L$ is long enough, with the increase of laser intensity two-photon absorption (TPA) will reach the curve saturation regime in an intensity range before the damage threshold. The occurrence of curve saturation means that the TPA process enters a specific physical regime of significant photon depletion, leading to the decline of modulation amplitude of anisotropic TPA. It seems that $L$ plays an important role in determining the laser intensity range for TPA entering the physical regime of photon depletion. In contrast, if $L$ is shortened to a sufficiently short length, the laser intensity for TPA entering the physical regime of photon depletion may be possible to exceed the laser damage threshold. That is, such an interaction length condition can prevent TPA from entering the photon depletion regime before the damage threshold. Below, we will discuss in detail the impact of $L$ on the TPA process based on the basic TPA formula of Z-Scan scheme.

In details, considering the alone mechanism of TPA occurring in a single-beam transmission measurement of (100) plane of zincblende crystal based on the Z-Scan technology, the transmitted beam intensity $I_t$ through the crystal with a length of $L$, that is, the laser-matter interaction length, is expressed as [17, 18]

$$I_t(r,z,\theta;t) = \frac{I(r,z;t)}{1+\beta(\theta)I(r,z;t)L}. \tag{1}$$

Here the linear absorption term has been omitted for the material with a zero linear absorption coefficient (like ZnSe and GaP at the 800-nm laser wavelength), $I$ is the incident beam intensity, and $\beta$ is the anisotropic TPA coefficient. From Equation (1), the expression for the normalized transmittance $T_l$ can be derived as

$$T_l(r,z,\theta;t) = \frac{1}{1+\beta(\theta)I(r,z;t)L}. \tag{2}$$

Then, with the relationship of $A = 1-T$ for absorptivity, from Equation (2) the modulation amplitude of asymmetric TPA can be derived [16]

$$\Delta A = A_{45} - A_0$$
$$= \frac{\sigma\beta_0 I(r,z;t)L/2}{1+(\beta_{45}+\beta_0)I(r,z;t)L+\beta_{45}\beta_0 I^2(r,z;t)L^2}. \tag{3}$$

Where $\sigma$ is the anisotropy coefficient, and '0' and '45' in the subscripts of some variables label the 0° and 45° crystal orientation angle, respectively.

For $\beta I(r,z;t)L \gg 1$ ($\beta$ represents $\beta(\theta)$, including the two specific angle cases of $\beta_0$ and $\beta_{45}$; the same below), Equation (3) can be approximated as [16]

$$\Delta A \approx \sigma/(2\beta_{45}I(r,z;t)L) \approx 0. \tag{4}$$

That is, when $I(r,z;t) \gg 1/(\beta L)$, the TPA process enters the photon depletion regime, and $\Delta A$ approaches 0. Therefore, for a specific $\beta$, smaller laser-matter interaction length $L$ will require higher laser intensity $I$ to meet the above condition for photon depletion.

For $\beta I(r,z;t)L \ll 1$, Equation (3) can be approximated as [16]

$$\Delta A \approx \sigma\beta_0 I(r,z;t)L/2 \approx (\beta_{45}-\beta_0)I(r,z;t)L. \tag{5}$$

Because $\beta_{45}-\beta_0$ is a slowly varying function of laser intensity $I$, when $I(r,z;t) \ll 1/(\beta L)$, an approximate linear relationship can be obtained as $\Delta A \propto I(r,z;t)$. Therefore, for a specific $\beta$, smaller laser-matter interaction length $L$ will favour higher laser intensity $I$ to meet the above condition for linear increasing trend.

Equations (4) and (5) mean that the appearance of the photon depletion regime will cause the modulation amplitude of orientation-dependent nonlinear absorption to reach a maximum value, and then decrease towards zero. In fact, as Equations (4) and (5) indicated, theoretically as long as the laser-matter interaction length gradually decreases, the photon depletion regime will move towards a higher intensity range, and the maximum modulation amplitude will also appear in a higher intensity range. Therefore, when the laser-matter interaction length is small enough, the laser intensity range for laser-matter interaction reaching the photon depletion regime may exceed the damage threshold. It means that under such a condition, the modulation amplitude of orientation-dependent nonlinear absorption will not saturate before the damage threshold. It is worth noting that then the dominant laser intensity ranges corresponding to the strongest modulation ranges of the orientation-dependent nonlinear absorption and the orientation-dependent laser-induced damage may become consistent, and thus the incompatibility discussed above will be solved.

In the single-beam transmission measurement, like the Z-scan measurement, fs laser pulses will propagate through the crystal. In this case, the laser-matter interaction length is determined by the crystal thickness. Typically, for the 1-mm-thick ZnSe and GaP crystal samples used in our previous research [16], when the fs laser intensity is around $10^{11}$ W/cm$^2$, photon depletion already becomes significant, and photons are mostly depleted during the transmission. Therefore, the periodic modulation of orientation-dependent transmission light will enter the weakened regime. In this study, for probing into the anisotropic third-order optical nonlinearity approaching the damage threshold in the physical regime without significant photon depletion, such as TPA and two-beam coupling (TBC) [19-22] in ZnSe and GaP irradiated by 800-nm fs laser at near-damage-threshold intensity, we use the pump-probe measurement scheme to replace the single-beam transmission measurement scheme, to achieve the significant reduction of the laser-matter interaction length. Based on the experimental scheme of non-collinear pump-probe measurement, the area of laser-matter interaction can be limited to the area of the pump and probe beams overlapping. Therefore, in the pump-probe experimental setup, the laser-matter interaction length can be significantly shorter than the crystal thickness, and thus the intensity dependence of the orientation-dependent third-order optical nonlinearity of ZnSe and GaP can be investigated in the near-damage-threshold intensity regime without significant photon depletion. On the other hand, the pump-probe experimental scheme also enables us to obtain the ultrafast dynamic characteristics of the orientation-dependent third-order optical nonlinearity of ZnSe and GaP. Actually, current studies on crystal orientation-dependent phenomena are mainly limited to steady-state experimental measurements, and thus the transient experimental measurements on these research topics, which would increase the understanding of the crystal orientation-dependent phenomena, is still lacking.

In summary, with regard to the various optical nonlinear phenomena, which always present strong crystallographic dependence under intense fs-laser irradiation, it is worth noting that the modulation amplitudes of anisotropic optical nonlinearities may present completely different trends as laser intensity approaches the damage threshold—it will decay or grow, seemingly contradictory to each other. Considering the similar physical origins for these crystallographic dependence phenomena—nonlinear absorption (ionization) under strong laser fields, the theory of TPA in third-order optical nonlinearities near the damage threshold, which presents the simplest theoretical scenario for nonlinear absorption, is explored to resolve this contradiction. It turns out that the laser intensity is not the alone physical quantity that determines the crystal orientation-dependent phenomena—the laser-matter interaction length is another important physical quantity that can significantly influence the phenomena. Based on the analytical analysis of the TPA process, it can be theoretically confirmed that the laser-matter interaction length is the key parameter to resolve the above-mentioned contradiction occurring in the crystallographic dependence phenomena, and sufficiently shortening the laser-matter interaction length can prevent nonlinear absorption from entering the photon depletion regime before the damage threshold, thus presenting a monotonically increasing intensity dependence for the crystallographic dependence phenomena. According to the theoretical derivation, we propose that, by introducing the fs-laser pump-probe measurement scheme, the laser-matter interaction length can be greatly reduced experimentally, and thus the probing of anisotropic optical nonlinearities in the near-damage-threshold intensity regime without significant photon depletion can be realized, which will display a non-saturation intensity dependence. In the following sections, we will conduct pump-probe experiments of anisotropic third-order optical nonlinearity near the damage threshold to confirm the above theoretical conclusions.

## 2. Experimental methods and setup

The experimental setup for the non-collinear pump-probe measurement is shown in Fig. 1. In the experiment, a femtosecond laser (Coherent, Legend Elite) with center wavelength of 800 nm, pulse duration of 70 fs, repetition frequency of 1 kHz was used as the light source. For the laser beam, an aperture (A1) was applied to modulate the beam diameter and make it suitable for the chopper. After passing through A1, the beam passed through a beam splitter (BS) and was divided into a probe beam and a pump beam, between which the delay was precisely controlled via a delay line (DL). The intensities of the pump and probe beams are modulated via neutral density filters (NDF1, NDF2). Because the experimental materials (ZnSe, GaP) can produce anisotropic third-order optical nonlinearity at very low laser intensity, in the pump-probe measurement the intensity of the probe beam was set to be a very low value of $\sim 2\times 10^8$ W/cm$^2$, to ensure that the probe beam could not cause observable nonlinear effects alone. With regard to the intensity of the pump beam, because this study focused on the crystallographic dependence of third-order optical nonlinearity with the laser intensity approaching the damage threshold, the intensity of the pump beam was set to be in the near-damage-threshold range of $10^{10}$-$10^{12}$ W/cm2, with the higher intensity being $\sim$ 80% of the damage threshold to ensure that there is no laser-induced damage on crystals during the long-time measurement. To modulate the polarization direction of the pump beam, a half-wave plate (HWP) was inserted in the pump path. Then, the pump and probe beams with an included angle of approximately 30°, were focused on the sample surface by plano-convex lenses (L1, L2) with focal lengths of 10 cm and 5 cm, respectively, forming a pump-probe overlapping area on the sample with a diameter less than 20 μm. Benefited from the small overlapping area and the non-collinear pump-probe setting, the laser-matter interaction length in the pump-probe measurement was less than 100 μm, far smaller than that of previous steady-state measurements [16] equal to the 1-mm thickness of the crystal plates. The sample placed in the center of an optic rotation mount with the surface perpendicular to the pump beam could be rotated around the optic axis of the pump beam with high accuracy. Thus, the relative angle between the crystal orientation and the pump and probe laser polarization directions could be continuously adjusted just by the rotation of the sample. To avoid the transmitted pump beam affecting the transmitted probe beam, the transmitted pump beam was blocked by a beam-block (BB). In addition, an aperture (A2) was also used to isolate the influence of pump beam and ambient light with a piece of shading cloth set from aperture to detector. After passing through A2, a focusing lens group (L3) is used to collect the probe beam signal. Finally, the received weak signal was amplified by a lock-in amplifier and then recorded by a computer.

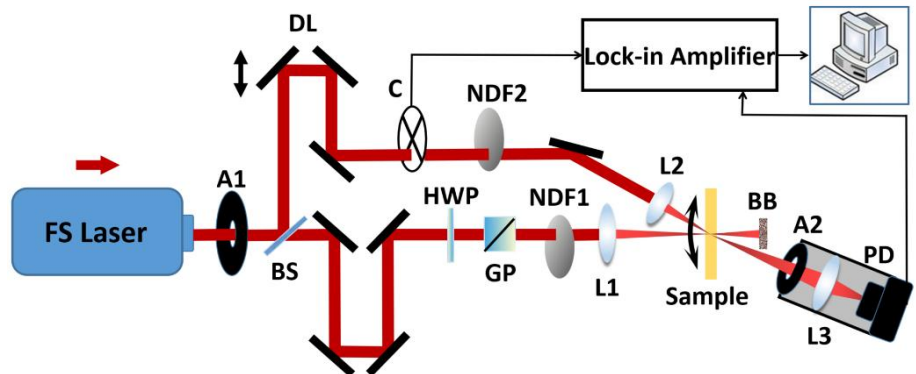

**Fig. 1.** Schematic diagram of the experimental setup for the non-collinear pump-probe measurement of anisotropic third-order optical nonlinearity. BS, Beam Splitter; A1-2, Aperture; DL, Delay Line; C, Chopper; HWP, Half Wave Plate; L1-3, Lens; NDF1-2, Neutral-Density Filter; PD, Electrophotonic Detector; BB, beam-block. The sample placed in the center of an optic rotation mount with the surface perpendicular to the pump beam.

The lattice structures of the narrow-bandgap crystals ([100]-cut ZnSe, [110]-cut ZnSe, [100]-cut GaP) used in this experiment are shown in Fig. 2. Both ZnSe and GaP crystals belong to the cubic crystal system (space group F-43m): ZnSe is a direct band gap material with a band gap of 2.8 eV; GaP is an indirect band gap material with an indirect band gap of 2.3 eV and a direct band gap of 2.9 eV [23]. The experimental crystals were purchased from two material suppliers of Hefei KMT Co., China and BJSCISTAR Co., China, which both had clearly marked the crystal planes and provided quality inspection reports for the crystal samples. Furthermore, we also carried out XRD crystal plane characterization on some samples and confirmed that the manufacturer's crystal plane markings were accurate. These 1-mm thick crystal plates are optically polished on both sides. In the study, the use of ZnSe and GaP is to enrich the material types of our experimental samples, and thus confirm the material universality of the above proposed mechanism that does not depend on specific values of intrinsic material parameters. In the measurements, two pump-probe polarization schemes of parallel polarization (polarization degeneracy) and orthogonal polarization (polarization non-degeneracy) are employed, which are suitable for the measurements of the TBC signal and the TPA signal, respectively. In the parallel polarization case (the polarization directions of the pump and probe beams are all set to be horizontal), around the time zero of the pump-probe interaction, besides the TPA signal, the probe beam may gain energy form the pump beam, or lose energy to the pump beam, and thus produce a bipolar-shape profile (the profile variation feature includes a peak and a valley, which is a very typical curve feature occurring in the fs TBC [19,20]) corresponding to the TBC signal in the transient curve, owing to the TBC energy transfer mechanism [19-21]. In the orthogonal polarization case (the polarization directions of the pump and probe beams are set to be vertical and horizontal, respectively), around the time zero of the pump-probe interaction, the TBC signal can be greatly eliminated due to the orthogonal polarization setting, left only the TPA signal located at the time zero of the transient curve.

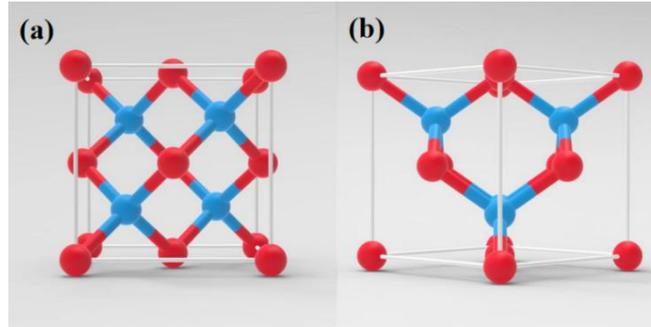

**Fig. 2.** Zinc blende lattice structure along the viewing directions perpendicular to the (a) (100) and (b) (110) crystal planes, respectively (Towards ZnSe and GaP: Red ball: Zn/Ga atom; Blue ball: Se/P atom).

Unlike traditional pump-probe measurement, here the pump-probe measurement has an additional control parameter of the crystal orientation angle, so the measured transient spectra can be presented in the form of transient 3D maps: the signal amplitude of the transmitted probe-beam intensity is presented as functions of the crystal orientation angle and the pump-probe delay. In the 3D maps, along the coordinate axis direction of delay time, the ultrafast dynamic characteristics of dominated dynamic mechanisms forms the basic morphological features of the 3D map, such as the bipolar profile of TBC, the valley profile of TPA, and the long-term evolution profile of free-carrier absorption (FCA); along the coordinate axis direction of crystal orientation angle, the orientation-dependent characteristics appear at the third-order nonlinear optical signal regions and cause the periodic modulation of the signals, such as the 4-fold modulation outline of the TPA and TBC regions for zinc blende crystals. In the pump-probe measurement process of the 3D map, for every 10 ° rotation of the sample, the transient spectrum corresponding to the crystal orientation angle is collected, and thus the entire data acquisition process should take several tens of minutes. Then, towards a specific transient signal found in the 3D map, we can keep the pump-probe delay line fixed at the specific delay time corresponding to the signal, and rotate the sample to measure the orientation-angle-dependent curve of the signal, characterizing the crystal orientation-dependent characteristics of the specific signal by slicing 3D images in a far shorter measure time than that of the 3D map.

## 3. Experimental results

### 3.1 The evolution of crystal orientation-dependent TBC signals

At first, the orientation-dependent pump-probe experiment is set to be the measurement state of polarization degeneracy—the polarization directions of the pump and probe beams are all horizontal. In Figs. 3(a), 3(b), and 3(c), the orientation-dependent transient spectrum of the (100) ZnSe crystal for different crystal orientations at a pump intensity of $5.73 \times 10^{11}$ W/cm$^2$ is shown as a transient 3D map towards different observation angles, which presents the signal amplitude of the probe-beam transmission intensity as functions of the crystal orientation angle and the pump-probe delay. Here, the polarization direction of the pump and probe beams oriented in the <001> crystal orientation is corresponding to 0 ° crystal orientation angle of the map. In Fig. 3(d), the transient spectrum curves of different pump intensities at 0 ° crystal orientation angle are demonstrated for examining the influence of pump intensity on transient spectral features.

From the map and curves shown in Fig. 3, it can be clearly seen that there is an obvious depression region constituting the basic contour feature of the ultrafast dynamic process near the zero delay time, which is originated from TPA (its dominated temporal position is labeled by the red arrow in Fig. 3(d)) of the probe beam induced by the pump beam, as well as FCA (its temporal range lasts beyond time zero as labeled by the green arrow in Fig. 3(d)) of the probe beam. Moreover, when the pump intensity is high enough, because of the wavelength and polarization degenerate pump-probe setting [19-21], the TBC (its dominated temporal position is labeled by the blue arrow in Fig. 3(d)) energy transfer will occurs between the pump and probe beams, which leads to a bipolar-shape profile formed in the transient (delay dependent) curve centered around the time zero. The entire energy transfer process lasts about 60fs (the entire TBC process is enlarged and highlighted by the blue square in Fig. 3(d)), which is consistent with the pulse width, and the central delay time of the energy transfer process for two beams is the time zero at which the two beams coincide. Thus, the basic dynamic process of the depression region is superimposed with the dynamic process of the bipolar region, forming the complete characteristics of the map and curves near the time zero. Note that, as the pump intensity increases, the delay time corresponding to the lowest point of the depression region gradually delays and deviates from the time zero, because the depression region comes from the superimposed contributions of TPA and FCA, and the FCA contribution will become more significant as the pump intensity increases, which are prevailed after the time zero for the dynamic process of free carrier generation mainly ascribed to TPA of the pump beam and electron impact ionization. As a result, the bipolar-shape profile tends to occur in the descending edge of the depression region, for which a narrow peak before time zero appears on the descending steep slope, while a narrow valley after time zero overlaps with the valley of the wide depression region in some extent.

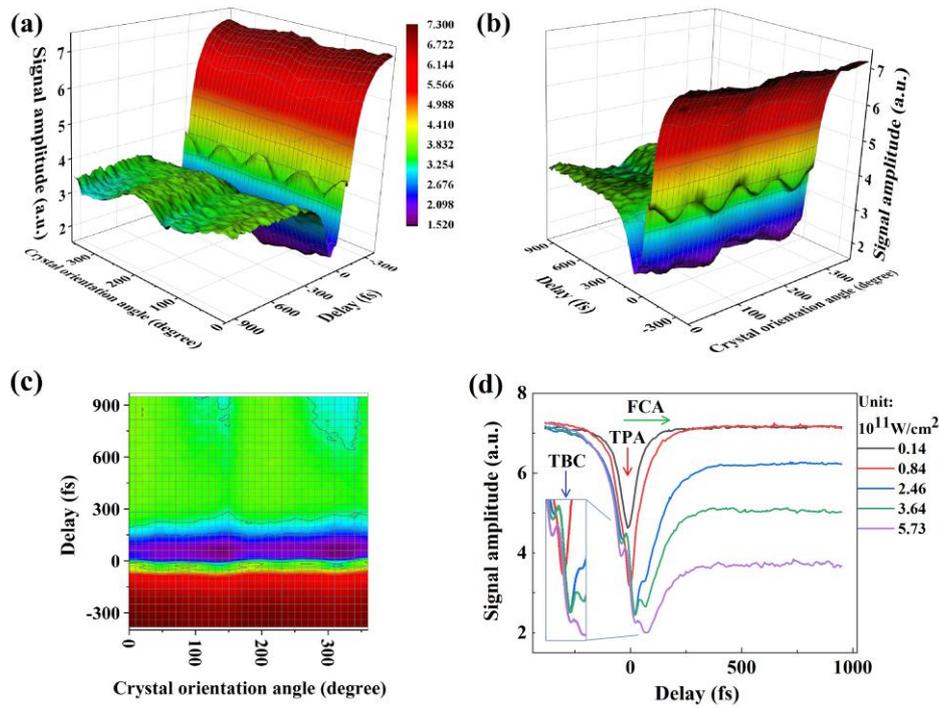

**Fig. 3.** The transient 3D map of the (100) ZnSe crystal at a pump intensity of $5.73 \times 10^{11}$ W/cm$^2$ for (a) and (b) the different side view angles, and (c) the top view angle. The polarization directions of the pump and probe beams are both horizontal, and the <001> crystal orientation oriented in the laser polarization direction is corresponding to 0° crystal orientation angle of the map. (d) The transient curves of different pump intensities at the 0° crystal orientation angle. The dominated temporal positions of TPA, FCA, and TBC are labeled by the red, green, and blue arrow, respectively.

In details, as shown in the 3D map of Figs. 3(a)-3(c) and the high-intensity curves of Fig. 3(d), the narrow peak for the probe beam obtaining energy from the pump beam in the TBC energy transfer process, appears slightly before the time zero. Interestingly, there is a periodic modulation for the peak value in the direction of the crystal orientation angle, leading to a uniform 4-fold modulation outline with a rotational period of 90°. Such a clear modulation indicates the TBC energy transfer between the pump and probe beams is also orientation-dependent, due to the third-order nonlinear optical essence of this phenomenon resembling to TPA of great orientation dependence [1,4,5,16]. In addition, the narrow valley located slightly after the time zero for the probe beam transferring energy to the pump beam in the TBC energy transfer process, also appears a periodic amplitude modulation. However, the small valley from TBC partially overlapping with the wide valley from TPA and FCA, results in superimposed asymmetric modulation features mainly introduced by FCA for the environmental-impact 2-fold amplitude fluctuations of free carrier generation, in particular the electron impact ionization, during long-time data acquisition of the 3D map (The data acquisition of a single 3D map takes tens of minutes, in which the measurement may undergo several cycles of laboratory temperature fluctuations of the ultra-clean room with constant temperature control of ±1 ℃ accuracy, such as two cycles in the crystal orientation angle direction for the map of Fig. 3).

Then, for comparison of the influence of crystallographic planes, similar pump-probe measurements on the (110) ZnSe crystal had also been carried out, as the transient 3D map and curves shown in Fig. 4. For the transient spectra of the (110) ZnSe crystal, besides the similar depression region around the time zero stemming from TPA and FCA, the bipolar-shape profile centered at the time zero stemming from TBC energy transfer may also occur in a similar way like that of the (100) ZnSe crystal. During the ultrafast TBC energy transfer process, as reflected by the similar positive and negative characteristics of bipolarity in Figs. 3(d) and 4(d), the probe light always first gains energy and then loses energy. Nevertheless, in the 3D map of Fig. 4, concerning the periodic amplitude modulation of the peak or valley of the TBC energy transfer process in the direction of crystal orientation angle, here the outline for (110) ZnSe provided with two strong primary peaks and two weak secondary peaks is significant different from that for (100) ZnSe provided with four uniform peaks, as shown in the map of Fig. 3.

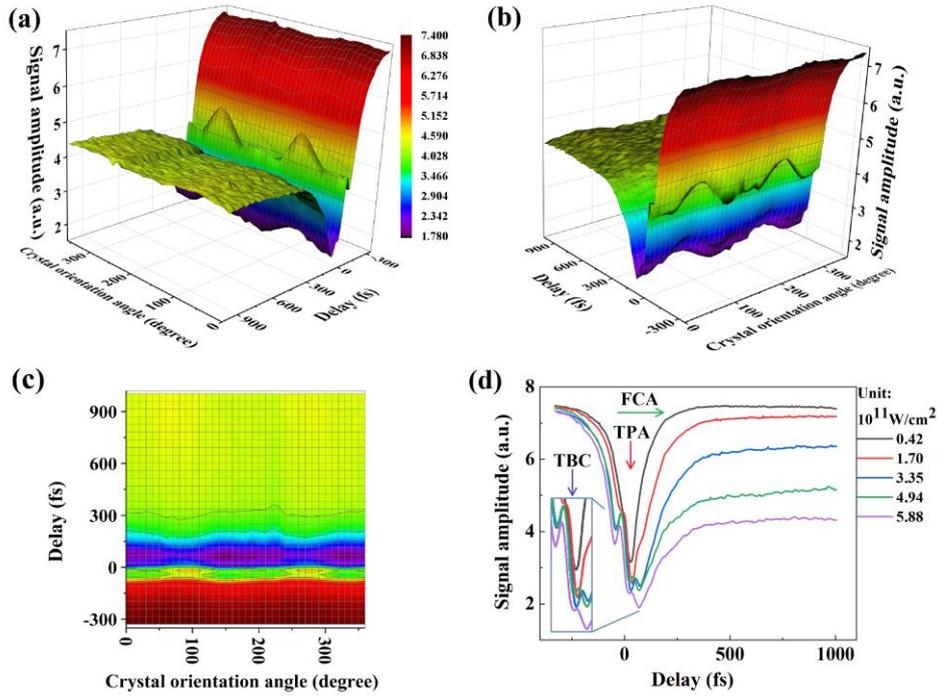

**Fig. 4.** The transient 3D map of the (110) ZnSe crystal at a pump intensity of $5.88 \times 10^{11}$ W/cm$^2$ for (a) and (b) the different side view angles, and (c) the top view angle. The polarization directions of the pump and probe beams are both horizontal, and the <011> crystal orientation oriented in the laser polarization direction is corresponding to 0° crystal orientation angle of the map. (d) The transient curves of different pump intensities at the 0° crystal orientation angle. The dominated temporal positions of TPA, FCA, and TBC are labeled by the red, green, and blue arrow, respectively.

In details, for the TBC outline of (110) ZnSe, although the two primary and two secondary peaks present alternating periodic distribution of rotated angles with 90° intervals, the four valleys of similar amplitudes present asymmetric distribution of rotated angles with alternate ~70° and ~110° intervals, like the steady-state orientation-dependent TPA curve of (110) ZnSe in the single-beam transmission measurement [16]. It can be seen that the directions of the modulation valleys are closely related to the ionic bond directions of the (110) crystal plane. That is, such an asymmetric 4-fold modulation of the outline agrees with the crystal lattice symmetry of zinc blende (110) plane (Fig 2(b)), which means that the asymmetric ionic bond distribution leads to the asymmetric distribution of the modulation valleys, resembling to the orientation-dependent origin of the steady-state TPA for (110) ZnSe. Actually, this similarity of orientation-dependent characteristics also occurs between the above TBC transient results and the TPA steady-state results [16] for (100) ZnSe. It seems that, because the same third-order nonlinear optical essence of TBC and TPA, their orientation-dependent measurements exhibits the same symmetric characteristics, no matter the transient measurement or the steady-state measurement.

On the other hand, here the two modulation outlines for the peak and valley values of the TBC energy transfer process have the same contour symmetry and variation characteristic, different from that in Fig. 3. This is because during the data acquisition of this 3D map the environmental-impact amplitude fluctuations of free carrier generation is much weaker than that in Figs. 3(a)-3(c), as shown by the different fluctuation characteristics of the free-carrier-evolution regions of two transient maps after time zero, and thus it will not significantly affect the contour features of the valley of the TBC energy transfer process in the 3D map of Fig. 4.

For further probing into the intensity dependence of the orientation-dependent third-order nonlinear optical effect, here we will focus on the TBC peak before the time zero, in order to obtain a direct analysis of the phenomenon originating from an alone mechanism. Specially, the evolution of crystal orientation-dependent outlines of the TBC peaks from the 3D maps corresponding to different pump intensities is presented in Fig. 5(a) towards the (100) ZnSe. It can be seem that the amplitude and regularity of the 4-fold modulation that is consistent with the 4-fold symmetry of the (100) surface of ZnSe crystal [16], still presents an upward trend with the increase of the pump intensity approaching the damage threshold, as shown in Figs. 5(b) and 5(c). Remarkably, with increasing light intensity towards the damage threshold, the increasing trend of modulation amplitude of orientation-dependent nonlinear optical signal is significantly different from the experimental results reported in previous studies [16], which always show a definite saturation or even attenuation trend.

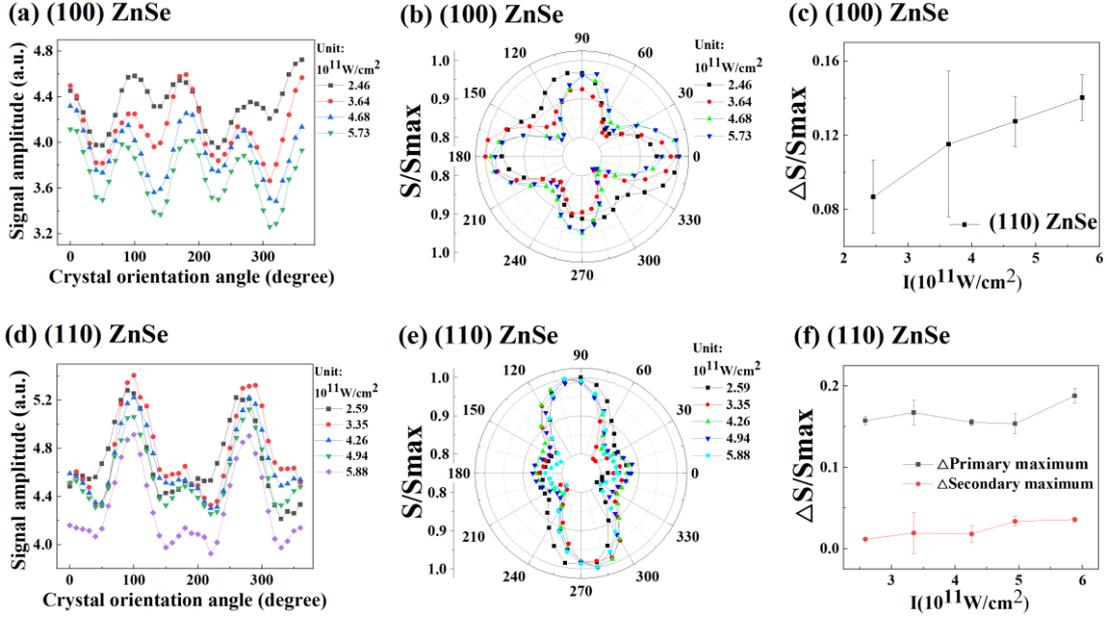

**Fig. 5.** The evolution of crystal orientation-dependent outlines of the TBC energy transfer peaks (delay time: ~20 fs before time zero) from the 3D maps of (100) and (110) ZnSe corresponding to different pump intensities with the parallel pump-probe polarization setting. (a) and (d), the orientation-dependent transmission curves from the 3D maps for a measured intensity series ; (b) and (e), the normalized modulation curves presented in polar coordinates; (c) and (f), the relationship between the pump intensity and the normalized modulation amplitude obtained by statistical averaging on the peak amplitude of the modulation curve.

Then towards the (110) ZnSe, concerning the intensity dependence of the modulation outlines, the evolution of crystal orientation-dependent outlines of the TBC peaks from the 3D maps corresponding to different pump intensities is presented in Fig. 5(d). From this modulation curve series, it seems that the amplitude of the asymmetric 4-fold modulation of the outline will not saturate as the pump intensity increases to near the damage threshold, as shown in Figs. 5(e) and 5(f), which agrees with the intensity-dependent trend of Fig. 5(c) and is completely different from the steady-state TPA results based on transmission measurements [16].

To better determine the exact intensity dependence of orientation-dependent TBC peaks, we maintain the pump-probe delay time at the peak position of the TBC energy transfer process, and then rotate the crystal to measure the crystal orientation-dependent transmission curves of the probe beam under a series of pump intensities, as shown in Fig. 6. This fixed delay time measurement can greatly shorten the data acquisition time and thus reduce the impact of environmental changes on the measurement. In Figs. 6(a) and 6(d), the orientation-dependent transmission curves of (100) and (110) ZnSe are presented for certain pump intensities of the measured series, which exhibit the same features as those shown in Fig. 5. Clearly, besides the significant orientation dependence corresponding to the crystallographic symmetries of (100) and (110) planes, these curves further confirm the increasing trend of the modulation amplitude with the increase of the pump intensity. As shown in Figs. 6(b) and 6(e), after normalizing the modulation curves and presenting them in polar coordinates, the increasing trend of modulation amplitude with the pump intensity is further manifested. Then, by performing statistical averaging on the peak amplitude of the modulation curve, we can obtain the relationship between the normalized modulation amplitude and the pump intensity for the TBC signals of (100) and (110) ZnSe, as shown in Figs. 6(c) and 6(f). Obviously, as the pump intensity approaches the damage threshold, the modulation amplitude of orientation-dependent TBC signals of (100) ZnSe, as well as that of (110) ZnSe mainly determined by the primary peak, presents a monotonous increasing trend, which is distinct from the decreasing trend in that of the steady-state TPA signals [16].

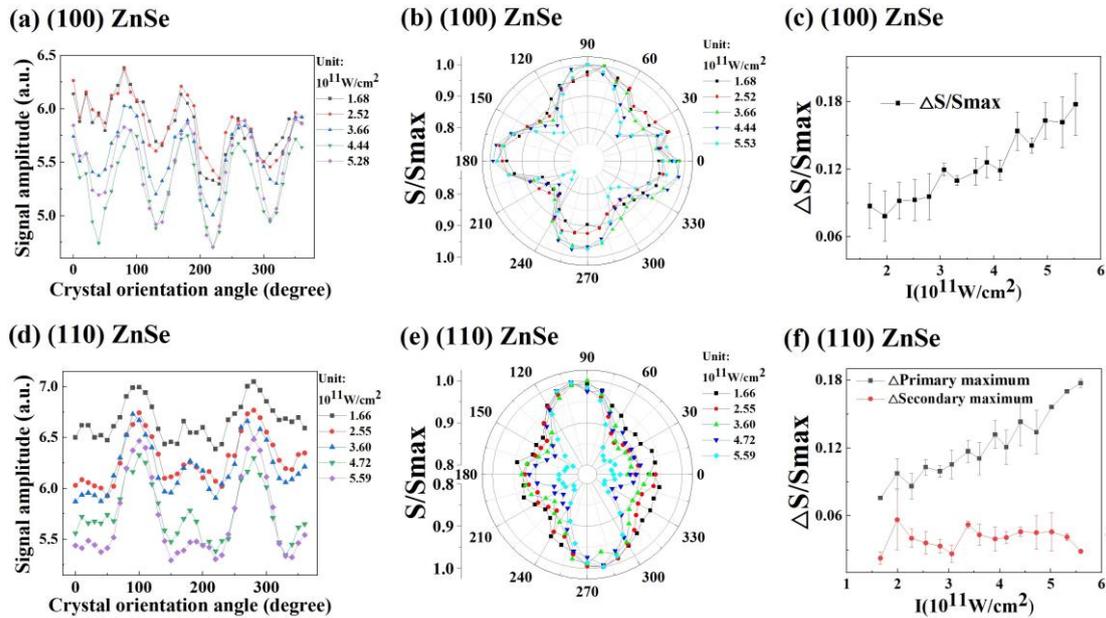

**Fig. 6.** Orientation dependent characteristics of the TBC signals of (100) and (110) ZnSe at the fixed pump-probe delay time corresponding to the TBC peak (delay time: ~20 fs before time zero) with the parallel pump-probe polarization setting. (a) and (d), the orientation-dependent transmission curves for certain pump intensities of the measured series ; (b) and (e), the normalized modulation curves presented in polar coordinates; (c) and (f), the relationship between the pump intensity and the normalized modulation amplitude obtained by statistical averaging on the peak amplitude of the modulation curve.

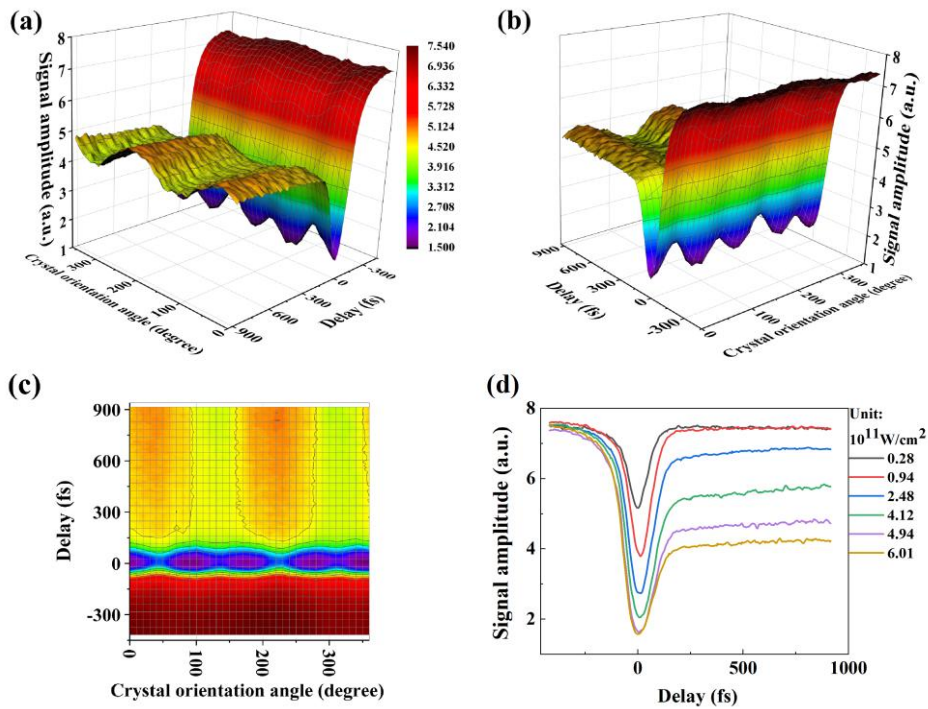

**Fig. 7.** The transient 3D map of the (100) ZnSe crystal at a pump intensity of $4.94 \times 10^{11}$ W/cm$^2$ for (a) and (b) the different side view angles, and (c) the top view angle. The polarization directions of the pump and probe beams are vertical and parallel to the horizontal plane, respectively, and the <001> crystal orientation oriented in the polarization direction of the pump beam is corresponding to 0 ° crystal orientation angle of the map. (d) The transient curves of different pump intensities at the 0 °crystal orientation angle.

*3.2 The evolution of crystal orientation-dependent TPA signals*

In Figs. 3 and 4, the TBC region and the TPA region partially overlap, especially in the TPA valley, so it is difficult to analyze the characteristics of TPA effects separately in the pump-probe measurement. Therefore, in order to better explore the crystal orientation dependence of TPA signals in the transient spectrum, we use a half-wave plate (HWP) to adjust the polarization direction of the pump beam to be vertical, which can greatly eliminate the TBC energy transfer between the pump and probe beams. After setting to the orthogonal polarization state, we repeated the above orientation-dependent pump-probe measurements of the (100) ZnSe and (110) ZnSe crystals, and obtained the transient 3D maps and curves without TBC signals, as shown in Figs. 7 and 8.

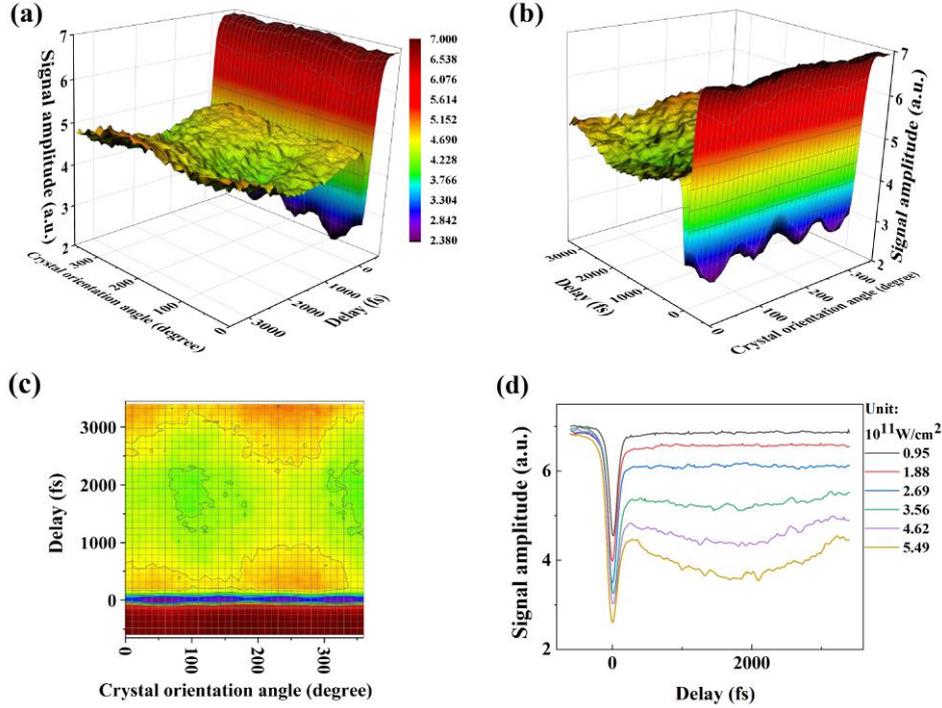

**Fig. 8.** The transient 3D map of the (110) ZnSe crystal at a pump intensity of $4.62 \times 10^{11}$ W/cm$^2$ for (a) and (b) the different side view angles, and (c) the top view angle. The polarization directions of the pump and probe beams are vertical and parallel to the horizontal plane, respectively, and the <111> crystal orientation oriented in the polarization direction of the pump beam is corresponding to 0 ° crystal orientation angle of the map. (d) The transient curves of different pump intensities at the 0 ° crystal orientation angle.

Herein, it can be clearly seen that the depression region corresponding to TPA of the probe beam induced by the pump beam significantly generates near the time zero in the maps. Due to the elimination of the TBC effects, the signal-to-noise ratio of TPA signals dominated at the time zero have been greatly improved. Therefore, the uniform 4-fold modulation and asymmetric 4-fold modulation at the bottom of the TPA valley near time zero can be observed clearly in the 3D maps of Figs. 7 and 8 for (100) and (110) ZnSe, respectively, indicating the strong orientation-dependent characteristic of TPA in the transient measurement, like that of TBC in the above measurements and that of TPA in the previous steady-state measurement [19,20]. On the other hand, due to the environmental impact during the long-time data acquisition of the maps, there also appear pronounced long-period modulations in the crystal orientation angle direction for the FCA regions after the time zero, like the contour features of the map in Fig. 3. Specially, in Fig. 8 the FCA region appears as an ultra-wide depression profile in the delay direction with the minimum occurring at the delay time of ~2000 fs, ascribed to the free carrier generation via electron impact ionization. Here the FCA regions after time zero exhibit no definite orientation-dependent features, which may be due to the small orientation-dependent modulation amplitude in the FCA region that is masked by the signal noises.

Then, Fig. 9 shows the evolution of crystal orientation-dependent outlines of the TPA valleys from the 3D maps corresponding to different pump intensities. Towards the curve series of (100) ZnSe in Fig. 9(a), the uniform 4-fold modulation of the TPA valleys shows similar features as that of the TBC peaks and the previous steady-state TPA results [16]. However, towards the curve series of (110) ZnSe in Fig. 9(d), the 4-fold modulation of the TPA valleys show significantly weakened asymmetric characteristics compared to that of the TBC peaks and the previous steady-state TPA results, which may be ascribed to the orthogonal pump-probe polarization setting that would weaken the non-orthogonal crystal orientation dependence effects.

Concerning the pump intensity dependence, the modulation amplitudes of these modulation curves do not show an observed increasing trend as the pump intensity increases, which seems to be inconsistent with the increasing trend of TBC results in Figs. 5 and 6. However, it can be seem that the average signal amplitude of the modulation curve decreases greatly with the increase of the pump intensity, which may suppress the increasing trend of modulation amplitude. Therefore, in order to more scientifically evaluate the dependence of modulation amplitude on light intensity, here it is very necessary to normalize the signal amplitude of the transmitted probe beam. As shown in Fig. 9(b) for (100) ZnSe with the normalized TPA curves presented in polar coordinates, after the normalization operation in general an increasing trend of normalized modulation amplitude can be figured out. Furthermore, by averaging the peak amplitude of the modulation curve, the relationship between the normalized modulation amplitude and the pump intensity can be obtained for the TPA signals of (100) ZnSe, as shown in Fig. 9(c). Clearly, for (100) ZnSe, as the pump intensity approaches the damage threshold, the normalized modulation amplitude still presents a general increasing trend, agreeing with the trend of the TBC signals in Figs. 5 and 6. However, for (110) ZnSe, due to the more significant signal noise, which may be introduced by the environmental impact during the long-term data acquisition, we cannot obtain a clear changing trend of the modulation amplitude with intensity, as shown in Fig. 9(e) and (f). It seems that the fixed-delay-time measurements are greatly needed to clarify this issue of (110) ZnSe.

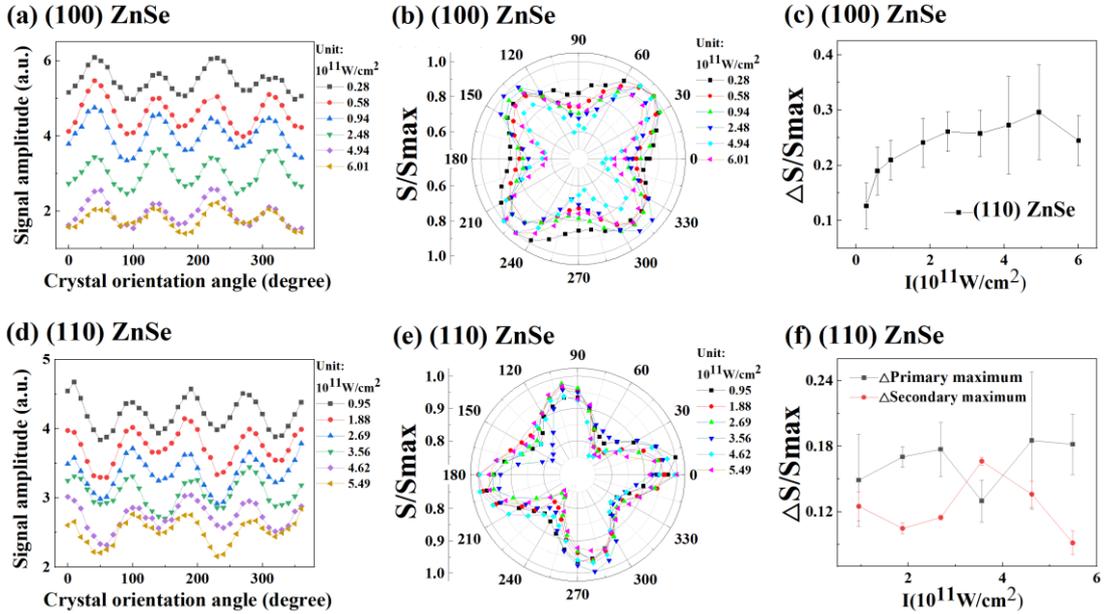

**Fig. 9.** The evolution of crystal orientation-dependent curves of the TPA valleys (delay time: ~0 fs, at time zero) from the 3D maps of (100) and (110) ZnSe corresponding to different pump intensities with the orthogonal pump-probe polarization setting. (a) and (d), the orientation-dependent transmission curves from the 3D maps for a measured intensity series ; (b) and (e), the normalized modulation curves presented in polar coordinates; (c) and (f), the relationship between the pump intensity and the normalized modulation amplitude obtained by statistical averaging on the peak amplitude of the modulation curve.

Therefore, resembling the measurements in Fig. 6, for the orthogonal pump-probe polarization setting, we also conducted measurements based on the fixed delay time way, to further eliminate the environmental impact caused by long-term data acquisition and probe into the orientation dependent characteristics of the transient signals under different delay time conditions, as shown in Figs. 10 and 11. In details, Figs. 10(a) and 11(a) show the orientation-dependent TPA curves of (100) and (110) ZnSe, respectively, at the pump-probe time zero for a pump intensity series approaching the damage threshold, which present similar features as that of Figs. 9(a) and 9(d) but with better modulation regularity and signal-to-noise ratio. Then, after the normalization operation, the normalized modulation curves presented in polar coordinates for (100) and (110) ZnSe are shown in Figs. 10(b) and 11(b), respectively, which both exhibit clear crystal orientation-dependent symmetry and obvious intensity-dependent modulation amplitude. It is worth noting that the fixed delay-time curves after the TPA valley, such as Figs. 10(c), 10(d), 11(c), and 11(d), do not exhibit observable crystallographic dependence. These curves collected in a short period of time only exhibit slight, random noise like data fluctuations with the varying of the crystal orientation angle, in stark contrast to the strong crystal orientation-dependent characteristic of the TPA valley region. Such definite curve features without periodic modulation further confirm that the long-period fluctuations occurring in the 3D maps of Figs.

3, 7, and 8 after the time zero come from the periodic changes in the environment during the long-term data acquisition process. On the other hand, the results indicate that the FCA process occurring after the TPA process in the pump-probe measurement of ZnSe does not exhibit observable crystallographic dependence characteristics at the noise level detected in our experiment. This is a slightly unexpected result, and its cause still needs further investigation.

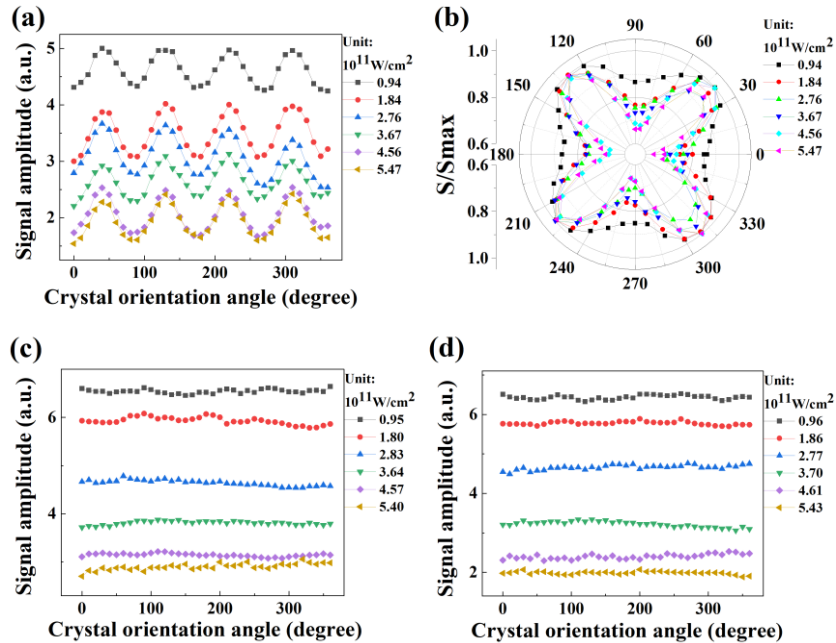

Fig. 10. Orientation dependent characteristics of the transient signals for (100) ZnSe under different delay time conditions with the orthogonal pump-probe polarization setting. (a) time zero (the bottom of TPA valley); (b) normalization of (a); (c) ~210 fs after time zero; (d) ~1960 fs after time zero (the bottom of FCA valley due to impact ionization). The <001> crystal orientation oriented in the polarization direction of the pump beam is corresponding to 0 ° crystal orientation angle.

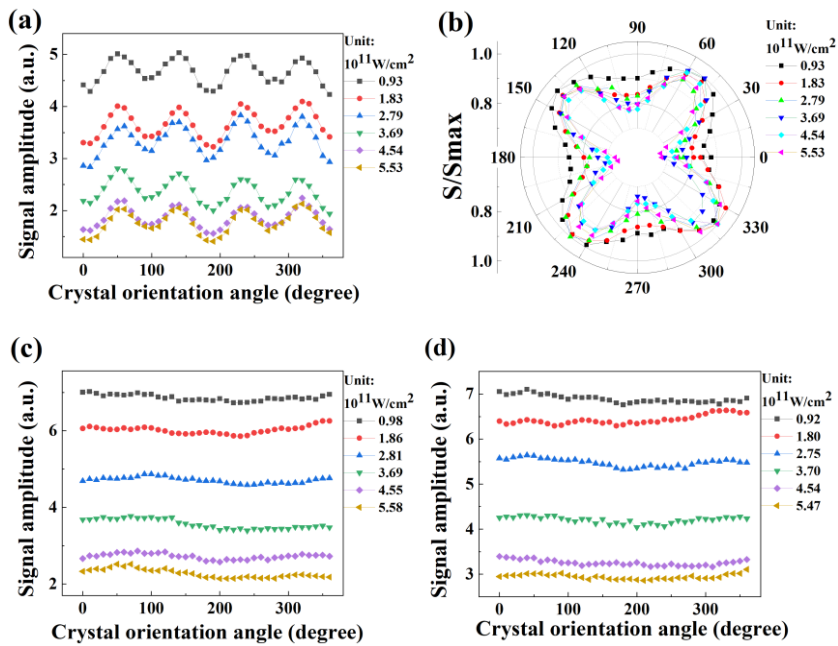

Fig. 11. Orientation dependent characteristics of the transient signals for (110) ZnSe under different delay time conditions with the orthogonal pump-probe polarization setting. (a) time zero (the bottom of TPA valley); (b) normalization of (a); (c) ~500 fs after time zero; (d) ~2000 fs after time zero (the bottom of FCA valley due to impact ionization). The <001> crystal orientation oriented in the polarization direction of the pump beam is corresponding to 0 ° crystal orientation angle.

In addition to (100) and (110) ZnSe, we also carried out the similar pump-probe measurements on (100) GaP, as shown in Fig. 12. Resembling to the results of (100) ZnSe, the orientation-dependent TPA curves of (100) GaP also show a clear 4-fold modulation and an increasing trend of the modulation amplitude. Actually, the modulation amplitude of the curves of (100) GaP is larger than that of (100) ZnSe, and the modulation regularity of the curves of (100) GaP is also better than that of (100) ZnSe. It turns out to be that for the TPA signals, GaP exhibits more pronounced crystallographic dependence characteristics than ZnSe, which may be related to the narrower bandgap of GaP. In addition, in Figs. 12(c) and 12(d), for (100) GaP the fixed delay-time curves after the TPA valley also do not exhibit observable crystallographic dependence, in agreement with the results of ZnSe.

Based on the orientation-dependent TPA curves of specific pump intensity series with good signal-to-noise ratio in Figs. 10, 11, and 12, we can obtain the definite relationship between the normalized modulation amplitude and the pump intensity for the TPA signals of (100) and (110) ZnSe, and (100) GaP, as shown in Fig. 13. It can be seen that all the intensity-dependent modulation amplitude curves demonstrate the same trend of data variation: overall, as the pump intensity increases, the normalized modulation amplitude of TPA signals shows an upward trend, which slows down with the pump intensity approaching the damage threshold. It should be noted that, although the modulation amplitude displays a preliminary tendency towards saturation towards the damage threshold, within the pump intensity range we measured, the orientation-dependent TPA signals has not reached the actual saturation range. Compared to the previous steady-state TPA results [16], for which at the intensity about $1.0 \times 10^{11}$ W/cm$^2$ the orientation-dependent TPA signals have reached complete saturation and entered the attenuation regime, here at 5 to 10 times of higher intensity the signals have not yet reached saturation. Our recent results confirm that the intensity saturation range of orientation-dependent TPA signals is directly related to the laser-matter interaction length: under the condition of a small enough laser-matter interaction length, the modulation amplitude of orientation-dependent TPA signals can continually increase to the damage threshold without saturation. Thus, under such a condition, the intensity regime of the orientation-dependent fs laser-induced TPA can overlap with the intensity regime of the orientation-dependent fs laser-induced damage, which could solve the problem of mismatch between the two intensity regimes as discussed in the introduction part. On the other hand, in Fig. 13 the upward trend of TPA shows a slowing-down trend with the pump intensity approaching the damage threshold, which may be due to the initial transition of the strong-field nonlinear optical regime from the multiphoton ionization regime to the tunneling ionization regime [24].

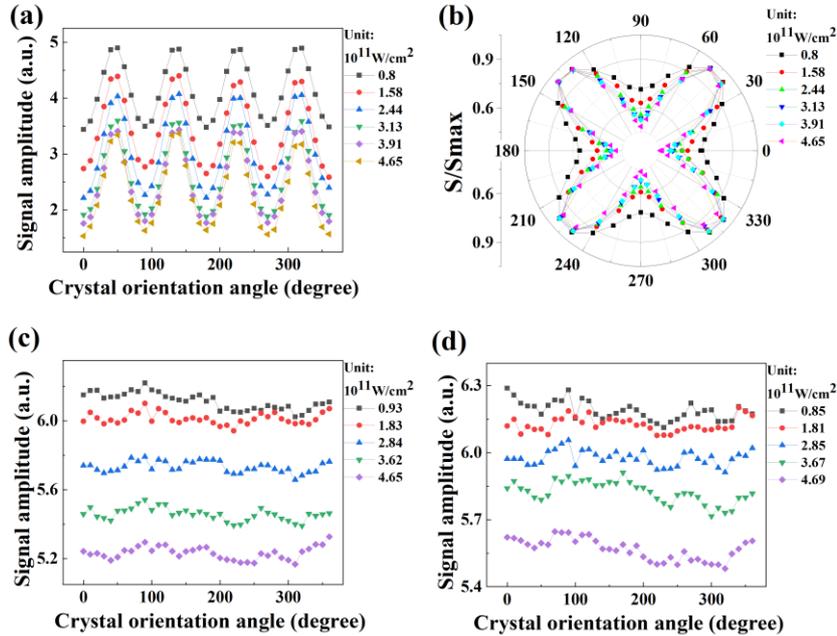

**Fig. 12.** Orientation dependent characteristics of the transient signals for (100) GaP under different delay time conditions with the orthogonal pump-probe polarization setting. (a) time zero (the bottom of TPA valley); (b) normalization of (a); (c) ~340 fs after time zero; (d) ~1480 fs after time zero (the bottom of FCA valley due to impact ionization). The <001> crystal orientation oriented in the polarization direction of the pump beam is corresponding to 0 ° crystal orientation angle.

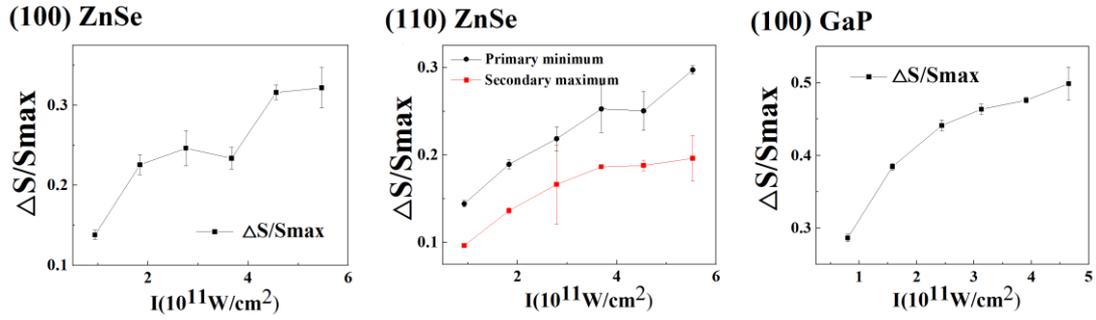

**Fig. 13.** The relationship between the normalized modulation amplitude and the pump intensity for the TPA signals of (a) (100) and (b) (110) ZnSe, and (c) (100) GaP.

## 4. Conclusion

In conclusion, in the study the intensity dependence of anisotropic third-order optical nonlinearity approaching the damage threshold in the narrow-bandgap crystals (ZnSe and GaP) is investigated via the femtosecond laser pump-probe measurements. Because the non-collinear pump-probe scheme can limit the area of laser-matter interaction to the localized area of the pump and probe beams overlapping, it can greatly reduce the laser-matter interaction length and thus realize the probing of orientation-dependent characteristics of nonlinear optical phenomena in the near-damage-threshold intensity regime without significant photon depletion. In the measured transient 3D map, the typical third-order nonlinear optical signals of TBC and TPA can be clearly found out, which both exhibit the pronounced orientation-dependent periodic modulation corresponding to a specific lattice symmetry. In particular, via the two pump-probe polarization schemes of parallel polarization and orthogonal polarization, the orientation-dependent characteristics of TBC and TPA can be independently detected, which show the similar uniform 4-fold and asymmetric 4-fold modulations corresponding to the lattice symmetries of (100) and (110) crystal planes, respectively, owing to the same third-order nonlinear optical essence of TBC and TPA.

Concerning the intensity dependence of the anisotropic TBC and TPA signals, although the symmetrical features of orientation-dependent curves of current transient measurements are consistent to that of previous steady-state measurements [16], there are significant difference on the intensity dependent characteristics between the two type measurements: as the laser intensity approaches the damage threshold, in previous steady-state measurements the modulation amplitude always show a definite saturation or even attenuation trend, whereas in current transient measurements the modulation amplitude still show a definite increasing trend, as confirmed by the fixed-delay-time measurements focusing on TBC and TPA. Such a definite upward trend of orientation-dependent third-order nonlinear optical effects in the near-damage-threshold regime, which has not been observed in previous studies, indicates that as long as the laser-matter interaction length is small enough, the third-order nonlinear optical phenomena can still be in a non-saturation physical regime till the damage threshold, and thus exhibit significant crystallographic dependence as that of laser-induced damage at the similar intensity ranges.

**Funding.** Natural Science Foundation of Guangdong Province (No. 2021A1515012335); National Natural Science Foundation of China (NSFC) (No.11274400); Pearl River S&T Nova Program of Guangzhou (201506010059); State Key Laboratory of High Field Laser Physics (Shanghai Institute of Optics and Fine Mechanics); State Key Laboratory of Optoelectronic Materials and Technologies (Sun Yat-Sen University).

**Acknowledgment.** The authors are grateful to Y F Liu and X R Zeng for his support in the experiments.

**Disclosures.** The authors declare no conflicts of interest.

**Data Availability.** Data underlying the results presented in this paper are not publicly available at this time but may be obtained from the authors upon reasonable request.